\documentclass[twocolumn]{aastex6}
\usepackage{rotating}
\usepackage{graphicx}
\usepackage{amssymb}
\usepackage{amsmath}
\usepackage{hyperref}


\def\gtrsim{\mathrel{\hbox{\rlap{\hbox{\lower4pt\hbox{$\sim$}}}\hbox{$>$}}}}
\def\lesssim{\mathrel{\hbox{\rlap{\hbox{\lower4pt\hbox{$\sim$}}}\hbox{$<$}}}}
\def\farcs{\hbox{$.\!\!^{\prime\prime}$}}

\begin{document}

\title{{\sl Chandra}, MDM, {\sl Swift}, and {\sl NuSTAR} observations confirming the  SFXT nature of  AX J1949.8+2534}
\author{Jeremy Hare\altaffilmark{1}, Jules P. Halpern\altaffilmark{2},  Ma\"{i}ca Clavel\altaffilmark{3}, Jonathan E. Grindlay\altaffilmark{4}, Farid Rahoui\altaffilmark{5}, and John A. Tomsick\altaffilmark{1} }
\altaffiltext{1}{Space Sciences Laboratory, 7 Gauss Way, University of California, Berkeley, CA 94720-7450, USA}
\altaffiltext{2}{Department of Astronomy, Columbia University, 550 West 120th Street, New York, NY 10027, USA}
\altaffiltext{3}{Universite Grenoble Alpes, CNRS, IPAG, F-38000 Grenoble, France}
\altaffiltext{4}{Harvard-Smithsonian Center for Astrophysics, 60 Garden Street, Cambridge, MA 02138, USA}
\altaffiltext{5}{Department of Astronomy, Harvard University, 60 Garden Street, Cambridge, MA 02138, USA}
\email{jhare@berkeley.edu}

\begin{abstract}
 AX J1949.8+2534 is a candidate supergiant fast X-ray transient (SFXT) observed in outburst by {\sl INTEGRAL}  (IGR J19498+2534). We report on the results of six {\sl Neil Gehrels Swift-XRT}, one {\sl Chandra}, and one {\sl NuSTAR} observation of the source. We find evidence of rapid X-ray variability on a few ks timescales. Fortunately, {\sl Chandra} observed the source in a relatively bright state, allowing us to confidently identify the optical/NIR counterpart of the source. We also obtained an optical spectrum of this counterpart, which shows an H$\alpha$ emission line and He I absorption features. The photometry and spectrum of the source allow us to constrain its distance, $\sim7-8$ kpc, and reddening, $A_V=8.5-9.5$. We find that the star is likely an early B-type Ia supergiant, confirming that  AX J1949.8+2534 is indeed  an SFXT. 
\end{abstract}

\section{Introduction}

The {\sl International Gamma-ray Astrophysics Laboratory} ({\sl INTEGRAL}) mission \citep{2003A&A...411L...1W} has been observing the hard X-ray sky for more than 16 years. During this time it has uncovered a large number of new Be high-mass X-ray binaries (BeHMXBs), while also dramatically increasing the population of supergiant HMXBs (sgHMXBs; see e.g., Figure 4 in \citealt{2015A&ARv..23....2W}). Additionally, {\sl INTEGRAL} has unveiled new sub-classes of sgHMXBs: supergiant fast X-ray transients (SFXTs; \citealt{2005A&A...444..221S,2006ESASP.604..165N}) and obscured supergiant HMXBs (e.g., IGR J16318-4848; \citealt{2003MNRAS.341L..13M}). There are also several sub-classes of SFXTs, which exhibit different dynamical ranges in their X-ray luminosities, from $\sim10^{2}-10^{5}$ and different flare time scales (see e.g., Table 2 in \citealt{2015A&ARv..23....2W,2011ASPC..447...29C}). Among all of these systems there are also  a number of ``intermediate'' HMXBs, that do not entirely fit into these categories and have properties that lie in between the classes (see e.g., \citealt{2018MNRAS.481.2779S}). In order to understand the different physical processes responsible for producing the emission observed in these sub-classes and how they depend on the binary parameters (e.g., orbital period, eccentricity), it is important to increase the number of known members by identifying them through their X-ray and optical properties.

Prior the to launch of {\sl INTEGRAL}, BeHMXBs dominated the population of known HMXBs \citep{2015A&ARv..23....2W}. BeHMXBs consist of a compact object (CO), typically a neutron star (NS), and a non-supergiant Be spectral type companion, generally having longer orbital periods, $P_{orb}\gtrsim20$ days, than sgHMXBs (see \citealt{2011Ap&SS.332....1R} for a review). The Be companions are rapidly rotating, have H$\alpha$ emission lines, and have an infrared (IR) excess compared to B stars \citep{2003PASP..115.1153P}. The H$\alpha$ emission and IR excess come from a dense equatorial decretion disk rotating at or near Keplerian velocity, likely formed due to the fast rotation of the star (see \citealt{2013A&ARv..21...69R} for a review). These systems can be both persistent, with X-ray luminosities on the order of $L_X\approx10^{34}-10^{35}$ erg s$^{-1}$, or transient \citep{2011Ap&SS.332....1R}. The transients become X-ray bright when the NS accretes material as it passes near (or through) the Be star's decretion disk. These systems exhibit two types of X-ray flares. Type I flares occur periodically or quasi-periodically when the NS passes through periastron, increasing the X-ray flux by an order of magnitude, and lasting a fraction of the orbital period ($\sim0.2\times P_{orb}\approx1-10$ days; \citealt{2011Ap&SS.332....1R}). Type II flares can occur at any orbital phase, reaching Eddington luminosity, and lasting a large fraction of an orbital period ($\sim10-100$ days; \citealt{2011Ap&SS.332....1R}). 

The sgHMXBs are composed of a CO and a supergiant O or B spectral type companion, typically with shorter orbital periods, $P_{orb}\lesssim10$ days, than BeHMXBs \citep{2011Ap&SS.332....1R,2015A&ARv..23....2W}.  If the CO orbits close enough to the companion star it can accrete via Roche lobe overflow, reaching X-ray luminosities up to $L_X\sim10^{38}$ erg s$^{-1}$ during an accretion episode \citep{2008A&A...484..783C,2015arXiv151007681C}. For longer period systems, the CO accretes from the fast ($\sim1000$ km s$^{-1}$) radiative wind of the supergiant companion, leading to persistent  X-ray luminosities of $L_X\sim10^{35}-10^{36}$ erg s$^{-1}$ \citep{2008A&A...484..783C}. These wind-fed systems are also often highly obscured ($N_H\gtrsim10^{23}$ cm$^{-2}$) by the wind of the companion, and in some cases, by an envelope of gas and dust around the entire binary system (see e.g., IGR J16318-4848; \citealt{2003A&A...411L.427W,2004ApJ...616..469F}). The supergiant stars in these systems also exhibit an H$\alpha$ emission line due to their winds, which is often variable in shape and intensity and can have a P-Cygni profile (see e.g., Vela X-1, IGR J11215-5952; \citealt{2001A&A...377..925B,2010ASPC..422..259L}).

{\sl INTEGRAL}'s wide field-of-view has enabled great progress in the study of HMXBs by discovering many SFXTs (see e.g., \citealt{2006ESASP.604..165N}, or \citealt{2013arXiv1301.7574S,2017mbhe.confE..52S} for reviews). Unlike the typical sgHMXBs, SFXTs exhibit much lower quiescent X-ray luminosities (as low as $L_X\sim10^{32}$ erg s$^{-1}$) with highly energetic ($L_X\sim10^{36}-10^{38}$  erg s$^{-1}$) X-ray flares lasting $\sim100-10,000$ s (see e.g., \citealt{2005A&A...441L...1I,2014A&A...568A..55R,2015A&A...576L...4R}). {\sl INTEGRAL} has also uncovered several systems, with flare to quiescent X-ray luminosity ratios of $10^{2}-10^{3}$,  lasting a few hours to days (see e.g., IGR J17354-3255; \citealt{2011MNRAS.417..573S,2013A&A...556A..72D}). These systems have larger variability than seen in classic sgHMXB systems, but longer variability timescales than in SFXTs, and therefore have been called intermediate SFXTs  \citep{2011MNRAS.417..573S,2011ASPC..447...29C}.

Several models have been put forward to describe the flaring phenomenon observed in SFXTs.  One possibility is that the flares are caused by the accretion of an inhomogeneous clumpy wind, produced by the high-mass stellar companion, onto the compact object (see e.g., \citealt{2005A&A...441L...1I,2006ESASP.604..165N,2007A&A...476..335W,2009MNRAS.398.2152D,2016A&A...589A.102B}). However, given the low-quiescent luminosities of some SFXTs, an additional mechanism for inhibiting the accretion onto the compact object is likely necessary (e.g., magnetic gating or sub-sonic accretion;  \citealt{2007AstL...33..149G,2008ApJ...683.1031B,2012MNRAS.420..216S}).

The source AX J1949.8+2534 was discovered by the {\sl ASCA} Galactic plane survey having an absorbed flux of 6$\times10^{-12}$ erg cm$^{-2}$ s$^{-1}$ in the 2-10 keV band \citep{2001ApJS..134...77S}. AX J1949.8+2534 ( AX J1949 hereafter) was then detected by {\sl INTEGRAL}  for the first time in the hard X-ray band during two short flaring periods in 2015/2016 \citep{2015ATel.8250....1S}\footnote{ The source is also referred to as IGR J19498+2534 in the {\sl INTEGRAL}  catalog of \cite{2017MNRAS.470..512K}.}. The first flaring episode lasted $\lesssim 1.5$ days and reached a peak flux  $F_X=1.1\times10^{-10}$ erg cm$^{-2}$ s$^{-1}$ in the 22-60 keV band, while the second flaring episode lasted $\sim4$ days with a similar peak flux $F_X= 1.0\times10^{-10}$ erg cm$^{-2}$ s$^{-1}$ in the 22-60 keV band \citep{2017MNRAS.469.3901S}. However, during the second flaring episode shorter time scale variability of $\sim$2-8 ks was detected, reaching a peak flux of 2$\times10^{-9}$ erg cm$^{-2}$ s$^{-1}$ in the 200 s binned light curve \citep{2017MNRAS.469.3901S}. The dynamic range of the flare to quiescent X-ray luminosities in the 20-40 keV band is $\gtrsim625$ \citep{2017MNRAS.469.3901S}.

In soft X-rays, \cite{2017MNRAS.469.3901S} also reported on a {\sl Neil Gehrels Swift-XRT} observation of the source, which was detected with an absorbed flux of 1.8$\times10^{-12}$ erg cm$^{-2}$ s$^{-1}$. This observation also provided a more accurate position for the source. However, there were two bright potential NIR counterparts within the 95\% {\sl Swift} positional uncertainty. Based on photometry, \cite{2017MNRAS.469.3901S} classified these two bright NIR sources as B0V and B0.5Ia spectral type stars, respectively, leading to the conclusion that this source was either a BeHMXB or SFXT type source. The bright flares observed from  AX J1949 disfavored the BeHXMB scenario, therefore, \cite{2017MNRAS.469.3901S} favored the SFXT interpretation.

In this paper we report on {\sl Neil Gehrels Swift-XRT}, {\sl Chandra}, and {\sl NuSTAR} legacy observations of the SFXT candidate  AX J1949. These observations are part of an ongoing program to identify the nature of unidentified {\sl INTEGRAL} sources by characterizing their broad-band (0.3-79.0 keV) X-ray spectrum. Additionally, we use the precise X-ray localization to identify the multi-wavelength counterpart for optical spectroscopic follow-up, which is also reported here.

\section{Observations and Data Reduction}
\label{obs_and_dat}
\subsection{{\sl Neil Gehrels Swift X-ray Observatory}}
The {\sl Neil Gehrels Swift} satellite's \citep{2004ApJ...611.1005G} X-ray telescope (XRT; \citealt{2005SSRv..120..165B}) has observed  AX J1949 a total of six times, once in 2016 and five more times in a span of $\sim20$ days in 2018.  The 2016 observation was originally reported on by \citealt{2017MNRAS.469.3901S} (although for consistency we reanalyze it here), while the five observations from 2018 are reported on for the first time in this paper. The details of each {\sl Swift-XRT} observation can be found in Table \ref{swiftobs}. We refer to each observation by the associated number shown in Table \ref{swiftobs}. The data were reduced using HEASOFT version 6.25 and the 2018 July version of the {\sl Swift-XRT} Calibration Database (CALDB). The {\sl Swift-XRT} instrument was operated in photon counting mode and the event lists were made using the {\tt xrtpipeline} task. To estimate the count rates, source significances, and upper limits we used the {\tt uplimit} and source statistics tool ({\tt sosta}) in the XIMAGE package (version 4.5.1). The count rates and upper limits were also corrected for vignetting, dead time, the point-spread function (PSF), and bad pixels/columns that overlap with the source region. 

We extracted the source spectrum for each {\sl Swift} observation in which the source was detected with a signal-to-noise ratio (S/N) $>3$ using XSELECT (version 2.4). A new ancillary response file was constructed using the {\tt xrtmkarf} task using the exposure map created by the {\tt xrtpipeline} task, and the corresponding response file was retrieved from the CALDB. The observed count rates (with no corrections applied) of the source in these observations were between $\sim(8-12)\times10^{-3}$ cts s$^{-1}$, so we used circular extraction regions with radii of 12 or 15 pixels depending on the observed count rate (see Table 1 in \citealt{2009MNRAS.397.1177E}). Background spectra were extracted from a source free annulus centered on  AX J1949. The {\sl Swift} spectra were grouped to have at least one count per bin and fit using Cash statistics (C-stat; \citealt{1979ApJ...228..939C}).

In this paper, all spectra are fit using XSPEC (version 12.10.1; \citealt{1996ASPC..101...17A}). We used the Tuebingen-Boulder ISM absorption model ({\tt tbabs}) with the solar abundances of \cite{2000ApJ...542..914W}. All uncertainties reported in this paper (unless otherwise noted) are 1 $\sigma$.

\begin{table}[h]
\caption{{\sl Swift-XRT} Observations and signal-to-noise ratio of  AX J1949.} 
\label{swiftobs}
\begin{center}
\renewcommand{\tabcolsep}{0.17cm}
\begin{tabular}{lcccccc}
\tableline 
$\#$ & ObsID &   offset &	Start Time	&	Exp. & S/N \\
\tableline 
 & &   arcmin &	MJD 	& s \\
 \tableline 
S0$^{a}$ & 00034497001 &  2.422  & 57503.446 & 2932 & 5.3\\
S1 &00010382001 &  2.997 & 58159.464 & 4910 & 5.5   \\
S2 &00010382002 &  2.240 & 58166.446 & 3369 & 5.4 \\
S3 &00010382003 &  2.441 & 58170.346& 1568 &  2.2\\
S4 &00010382004 &  2.938 & 58173.336 & 4662 & 2.6 \\
S5 &00010382005 &  0.424 & 58179.584 & 5482 & 4.5 \\
\tableline 
\end{tabular} 
\tablenotetext{a}{ This observation was originally reported on by \cite{2017MNRAS.469.3901S}.}
\end{center}
\end{table}


\subsection{{\sl Chandra} X-ray Observatory}
We observed  AX J1949 using the {\sl Chandra} Advanced CCD Imaging Spectrometer (ACIS; \citealt{2003SPIE.4851...28G}) on 2018 February 25 (MJD 58174.547; obsID 20197) for 4.81 ks. The source was observed by the front-illuminated ACIS-S3 chip in timed exposure mode and the data were telemetered using the ``faint'' format. A 1/4 sub-array was used to reduce the frame time to 0.8 s, ensuring that the pileup remained $<3\%$ throughout the observation. All {\sl Chandra} data analysis was performed using the {\sl Chandra} Interactive Analysis of Observations (CIAO) software version 4.10 and CALDB version 4.8.1. The event file was reprocessed using the CIAO tool {\tt chandra\_repro} prior to analysis. 

To locate all sources in the field of view, the CIAO tool {\tt wavdetect} was run on the 0.5-8 keV band image\footnote{ We ran {\tt wavdetect} using the exposure map and PSF map produced by the CIAO tools {\tt fluximage} and {\tt mkpsfmap}, respectively.}. Only one source, the counterpart of  AX J1949, was detected at the position R.A.$=297.48099^{\circ}$, Dec.$=25.56639^{\circ}$ with a statistical 95\% confidence positional uncertainty of $0\farcs13$ estimated using the empirical relationship (i.e., equation 12) from \cite{2007ApJS..169..401K}. We were unable to correct for any systematic uncertainty in the absolute astrometry because only one X-ray source was detected. Therefore, we adopt the overall 90\% {\sl Chandra} systematic uncertainty of 0\farcs8\footnote{\url{http://cxc.harvard.edu/cal/ASPECT/celmon/}} and convert it to the 95\% uncertainty by multiplying by 2.0/1.7. The statistical and systematic errors were then added in quadrature, giving a 95\% confidence positional uncertainty radius of $0\farcs95$ for the source.

The {\sl Chandra} energy spectrum of  AX J1949 was extracted from a circular region centered on the source and having a radius of 2$''$, enclosing $\sim95\%$ of the PSF at 1.5 keV\footnote{See Chapter 4, Figure 4.6 at \url{http://cxc.harvard.edu/proposer/POG/html/}.}, and containing  260 net counts. The background spectrum was extracted from a source free annulus centered on the source. Given the small number of counts, we fit the {\sl Chandra} spectrum using Cash statistics (C-stat; \citealt{1979ApJ...228..939C}).

We have also extracted the {\sl Chandra} light curves using a number of different binnings to search for spin and orbital periods in the data. Prior to extraction, the event times were corrected to the Solar system barycenter using the CIAO tool {\tt axbary}.

\subsection{{\sl NuSTAR}}
The {\sl Nuclear Spectroscopic Telescope Array} ({\sl NuSTAR}; \citealt{2013ApJ...770..103H}) observed  AX J1949 on 2018 February, 24 (MJD 58173.070; obsID 30401002002) for 45 ks. We reduced the data using the NuSTAR Data Analysis Software (NuSTARDAS) version 1.8.0 with CALDB version 20181022. Additionally, we filtered the data for background flares caused by {\sl NuSTAR's} passage through the South Atlantic Anomaly (SAA) using the options {\tt saacalc}=2, {\tt saamode}=optimized, and {\tt tentacle}=yes, which reduced the total exposure time to 43 ks.

The source's energy spectra from the FPMA and FPMB detectors were extracted from circular regions with radii of 45$''$ centered on the source position. The background spectra were extracted from source-free regions away from  AX J1949, but on the same detector chip. The {\sl NuSTAR} spectra were grouped to have at least one count per bin. {\sl NuSTAR} light curves were also extracted  from both the FPMA and FPMB detectors in the 3-20 keV energy range using a number of different bin sizes (i.e., 100 s, 500 s, 1 ks, 5 ks). All {\sl NuSTAR} light curves plotted in this paper show the averaged (over the FPMA and FPMB detectors) net count rate.

\subsection{MDM Spectroscopy}

The accurate position of the X-ray counterpart to  AX J1949 provided by {\sl Chandra} has allowed us to identify the optical/NIR counterpart to the source. The optical/NIR counterpart is the brightest source (see Table \ref{mw_mag}) considered by \cite{2017MNRAS.469.3901S} as a potential counterpart (i.e., their source 5). This source has a {\sl Gaia} position  R.A.$=297.480949745(9)^{\circ}$ and Decl.$=25.56659555(1)^{\circ}$, which is $\sim0\farcs76$ away from the {\sl Chandra} source position and within the 2$\sigma$ {\sl Chandra} positional uncertainty.

\begin{figure*}
\centering
\includegraphics[trim={0 0 0 0},scale=0.60]{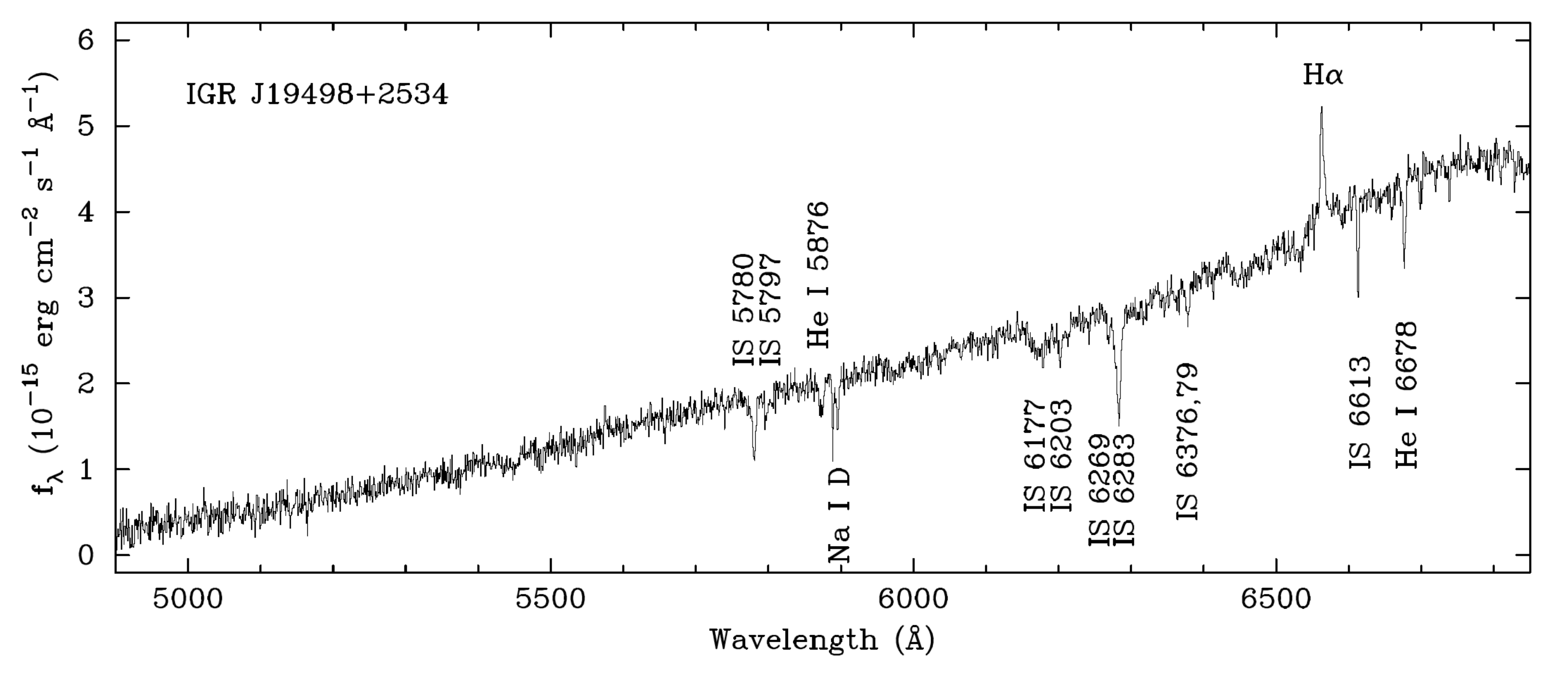}
\caption{MDM 2.4 m optical spectrum of  AX J1949 showing H$\alpha$ emission and He I absorption. The other spectral absorption features are diffuse interstellar bands caused by the large reddening towards the source ($A_V=$8.5-9.5, see Section \ref{spec_type}).  (Supplemental data for this figure are available in the online journal.)
\label{opt_spec}
}
\end{figure*}

On 2018 October 18, a 600~s spectrum was obtained with the Ohio State
Multi-Object Spectrograph on the 2.4~m Hiltner telescope of the
MDM Observatory on Kitt Peak, Arizona.  A $1.\!^{\prime\prime}2$
wide slit and a volume-phase holographic grism provided a dispersion
of 0.72 \AA\ pixel$^{-1}$ and a resolution of $\approx3$ \AA\ over
the wavelength range 3965--6878 \AA.  The reduced spectrum is
shown in Figure \ref{opt_spec}, where it can be seen that flux is not well
detected below 4900 \AA\ due to the large extinction to the star.
Although a standard star was used for flux calibration, the narrow
slit and partly cloudy conditions are not conducive to
absolute spectrophotometry. Therefore, as a last step we have 
scaled the flux to match the $V$ magnitude from Table \ref{mw_mag}.

\section{Results}

\subsection{X-ray spectroscopy}
\label{xspectro}
\begin{figure}
\centering
\includegraphics[trim={0 0 0 0},scale=0.38]{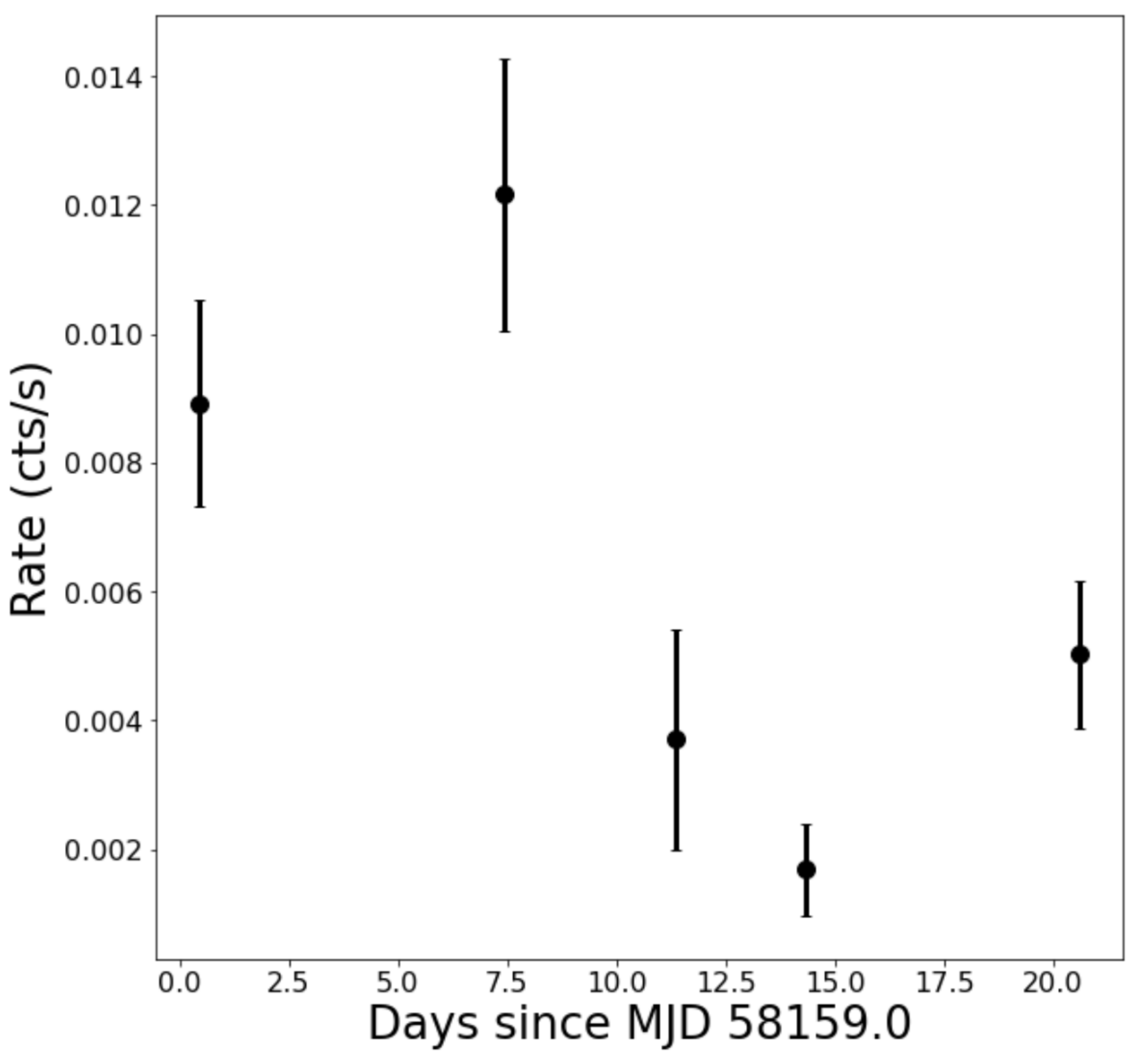}
\caption{{\sl Swift-XRT} net count rate light curve of  AX J1949 during the $\sim$20 day observing period in 2018, which included observations S1 through S5 (see Table \ref{swiftobs}). 
\label{xrt_cr}
}
\end{figure}

\begin{figure}
\centering
\includegraphics[trim={0 0 0 0},scale=0.41]{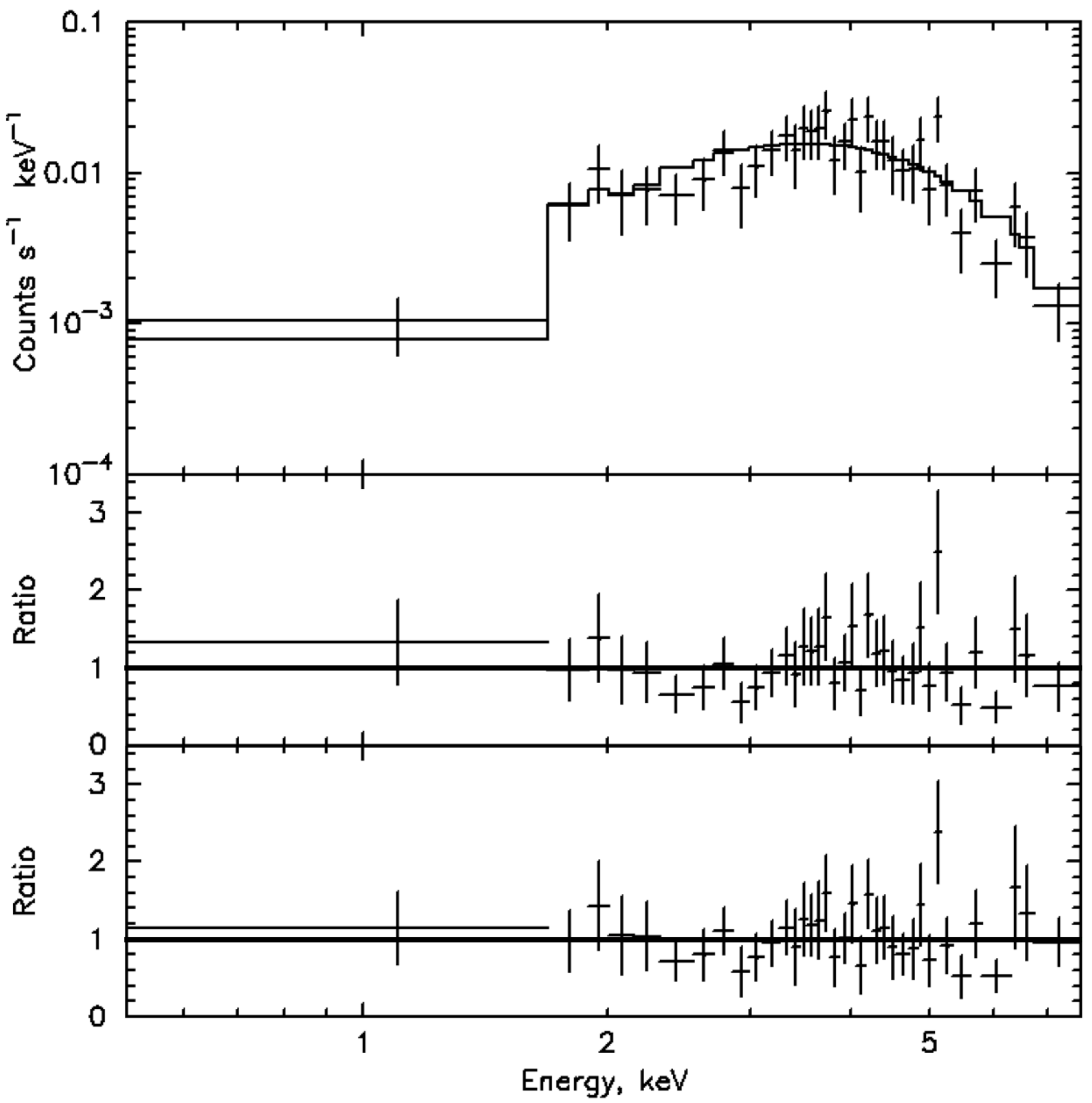}
\caption{{\sl Chandra} 0.5-8 keV spectrum with best-fit power-law model (see Section \ref{xspectro} for best-fit parameters). The middle panel shows the ratio of the data to the power-law model, while the bottom panel shows the ratio of the data to the blackbody model. The spectra were binned for visualization purposes only.
\label{cxo_spec}
}
\end{figure}

 AX J1949 is a variable X-ray source (see Figure \ref{xrt_cr}), but fortunately, {\sl Chandra} observed it during a relatively bright state, allowing for the most constraining X-ray spectrum of all of the observations reported here. We fit the {\sl Chandra} spectrum in the 0.5-8 keV range with two models, an absorbed power-law model and an absorbed blackbody model  (see Figure \ref{cxo_spec}). The best-fit power law model has a hydrogen absorption column density $N_{H}=7.5^{+1.6}_{-1.4}\times10^{22}$ cm$^{-2}$, photon index $\Gamma=1.4\pm0.4$, and absorbed flux $(2.0\pm0.3)\times10^{-12}$ erg cm$^{-2}$ s$^{-1}$ (in the 0.3-10 keV band) with a C-stat of 121.5 for 182 d.o.f. On the other hand, the best-fit blackbody model has hydrogen absorption column density $N_{H}=(4.7\pm1)\times10^{22}$ cm$^{-2}$, temperature $kT=1.5\pm0.2$ keV, radius $r_{\rm BB}=150^{+30}_{-20}d_{\rm 7kpc}$ m, and absorbed flux $(1.6\pm0.2)\times10^{-12}$ erg cm$^{-2}$ s$^{-1}$ (in the 0.3-10 keV band), where $d_{\rm 7kpc}=d/$(7 kpc) is the assumed distance to the source (see Section \ref{spec_type}), with a C-stat of 119.3 for 182 d.o.f. In either case, the photon-index, or the blackbody temperature and emitting radius, are consistent with those observed in SFXTs (see e.g., Table 4 in \citealt{2009MNRAS.399.2021R}).  Therefore, since both models fit the data about equally well we cannot distinguish between them.

We extracted the {\sl Swift-XRT} spectra for all observations where the source was detected with a S/N$>3$ (i.e., S0, S1, S2, and S5). The source is not bright enough in any of these single observations to fit a constraining spectrum. Therefore, we jointly fit the {\sl Chandra} spectrum with the four {\sl Swift-XRT} spectra, tying together the hydrogen absorbing column, but leaving the photon index and normalization free to vary for each spectrum. The best-fit parameters can be seen in Table \ref{spec_par}. There is marginal evidence ($\sim2\sigma$) of spectral softening in the spectrum of S2, which could also be due to a variable hydrogen absorbing column as is often seen in SFXTs (see e.g., \citealt{2009MNRAS.397.1528S,2016A&A...596A..16B,2018arXiv181111882G}). We have also fit the spectra with an absorbed power-law model after tying together the $N_H$ and photon-index before fitting. When this is done, the best fitting spectrum has a hydrogen absorption column density $N_H=5.6^{+1.1}_{-1.0}\times10^{22}$ cm$^{-2}$ and photon-index $\Gamma=1.2\pm0.3$ with a C-stat of 275.0 for 295 d.o.f.

Unfortunately,  AX J1949 was simultaneously observed with {\sl NuSTAR} during the S4 {\sl Swift-XRT} observation, when the source was at its faintest flux level (see Figure \ref{fluxes}).  Although the source was still jointly detected (i.e., FPMA+FPMB) with a significance of 5.3$\sigma$ and 5.7$\sigma$ (in the 3-79 keV and 3-20 keV energy bands, respectively), there were too few counts to constrain its spectrum. However, we extracted the 3$\sigma$ upper-limit flux in the 22-60 keV energy range for later comparison to {\sl INTEGRAL}. To do this, we calculated the 3$\sigma$ upper-limit on the net count rate of the source in the 22-60 keV energy range (0.0011 cts s$^{-1}$).  SFXTs often show a cutoff in their spectra between $\sim5-20$ keV (see e.g., \citealt{2011MNRAS.412L..30R,2017ApJ...838..133S}) so we use a power-law model with the best-fit absorption ($N_H=5.6\times10^{22}$ cm$^{-2}$) and a softer photon index ($\Gamma=2.5$) that is typical for SFXTs detected in hard X-rays by {\sl INTEGRAL} (see e.g., \citealt{2008A&A...487..619S,2011MNRAS.417..573S}) to convert the count rate to flux. The 3$\sigma$ upper-limit on the flux in the 22-60 keV energy range is $F=5.8\times10^{-13}$ erg cm$^{-2}$ s$^{-1}$.


\begin{table}[t!]
\caption{Photon indices for the simultaneous best-fit absorbed power-law model.} 
\label{spec_par}
\begin{center}
\renewcommand{\tabcolsep}{0.11cm}
\begin{tabular}{lccccc}
\tableline 
Obs. & $N_H$ &  $\Gamma$ &		C-stat	 &d.o.f. \\
\tableline 
 &   $10^{22}$ cm$^{-2}$ &	  	 &   \\
\tableline 
CXO & 5.6$^{+1.1}_{-1.0}$ & 1.0$\pm0.3$ &   243.4 & 286 \\ 
S0 & 5.6\tablenotemark{a} & 1.0$\pm0.5$ &  -- & -- \\
S1 & 5.6\tablenotemark{a} & 1.5$\pm0.6$ &  -- & -- \\
S2 & 5.6\tablenotemark{a} & 2.5$^{+0.6}_{-0.5}$ & -- & --\\
S5 & 5.6\tablenotemark{a} & 2.2$\pm0.7$ & -- & --\\
\tableline 
\end{tabular} 
\tablenotetext{a}{The hydrogen absorption column density ($N_{\rm H}$) values were tied together for the fit reported in this table.}
\end{center}
\end{table}

\subsection{X-ray variability and timing}
\begin{figure}
\centering
\includegraphics[trim={0 0 0 0},scale=0.36]{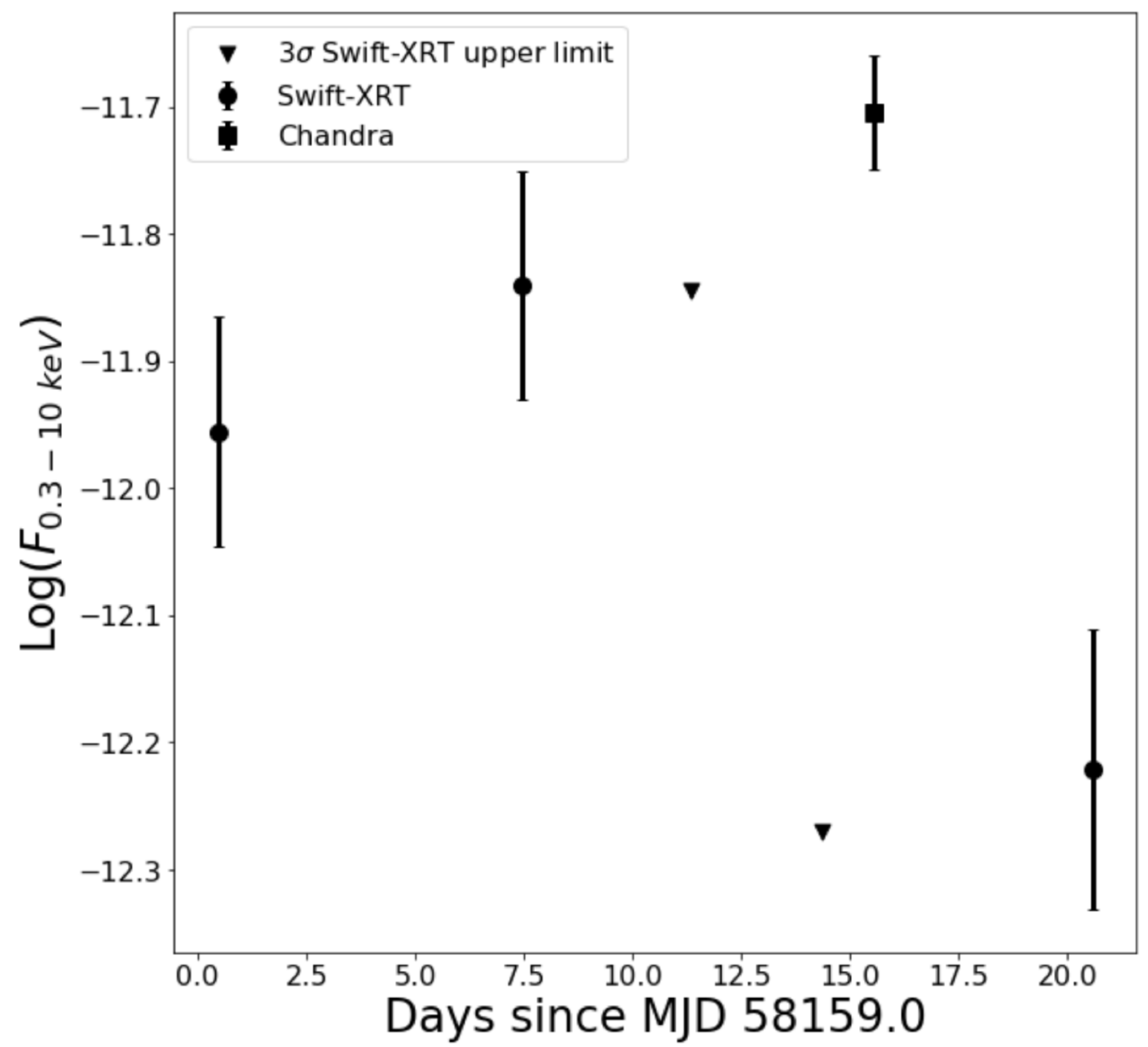}
\caption{Fluxes of  AX J1949 in the 0.3-10 keV energy range from the jointly fit {\sl Swift-XRT} and {\sl Chandra} (CXO) spectra (see Section \ref{xspectro}). The points show the fluxes from the different observations (circles for {\sl Swift-XRT} and squares for {\sl Chandra}), while the triangles show the 3$\sigma$ upper-limit of the source's flux when it was not detected in the {\sl Swift-XRT} observations. The flux for observation S0 is excluded. The {\sl NuSTAR} observation was concurrent with the S4 observation (i.e., the second triangle), when the flux of the source was at its minimum.
\label{fluxes}
}
\end{figure}

The fluxes of the source are not constrained if the photon-index is left free when jointly fitting the {\sl Chandra} and {\sl Swift-XRT} spectra. Therefore, to extract the fluxes from the {\sl Swift} and {\sl Chandra} observations we use the jointly fit model (i.e., $\Gamma=1.2$ and $N_H=5.6\times10^{22}$ cm$^{-2}$; see Section \ref{xspectro}). The fluxes derived from these fits can be seen in Figure \ref{fluxes}. Additionally, we assumed the same best-fit power-law model when converting the {\sl Swift} 3$\sigma$ upper-limit count rates to fluxes. The flux from observation S0 is not shown in this figure but is $F_{0.3-10 \ keV}=(1.4\pm0.3)\times10^{-12}$ erg cm$^{-2}$ s$^{-1}$, which is consistent within 2$\sigma$ of the value reported by \cite{2017MNRAS.469.3901S}.

\begin{figure}
\centering
\includegraphics[trim={0 0 0 0},scale=0.3]{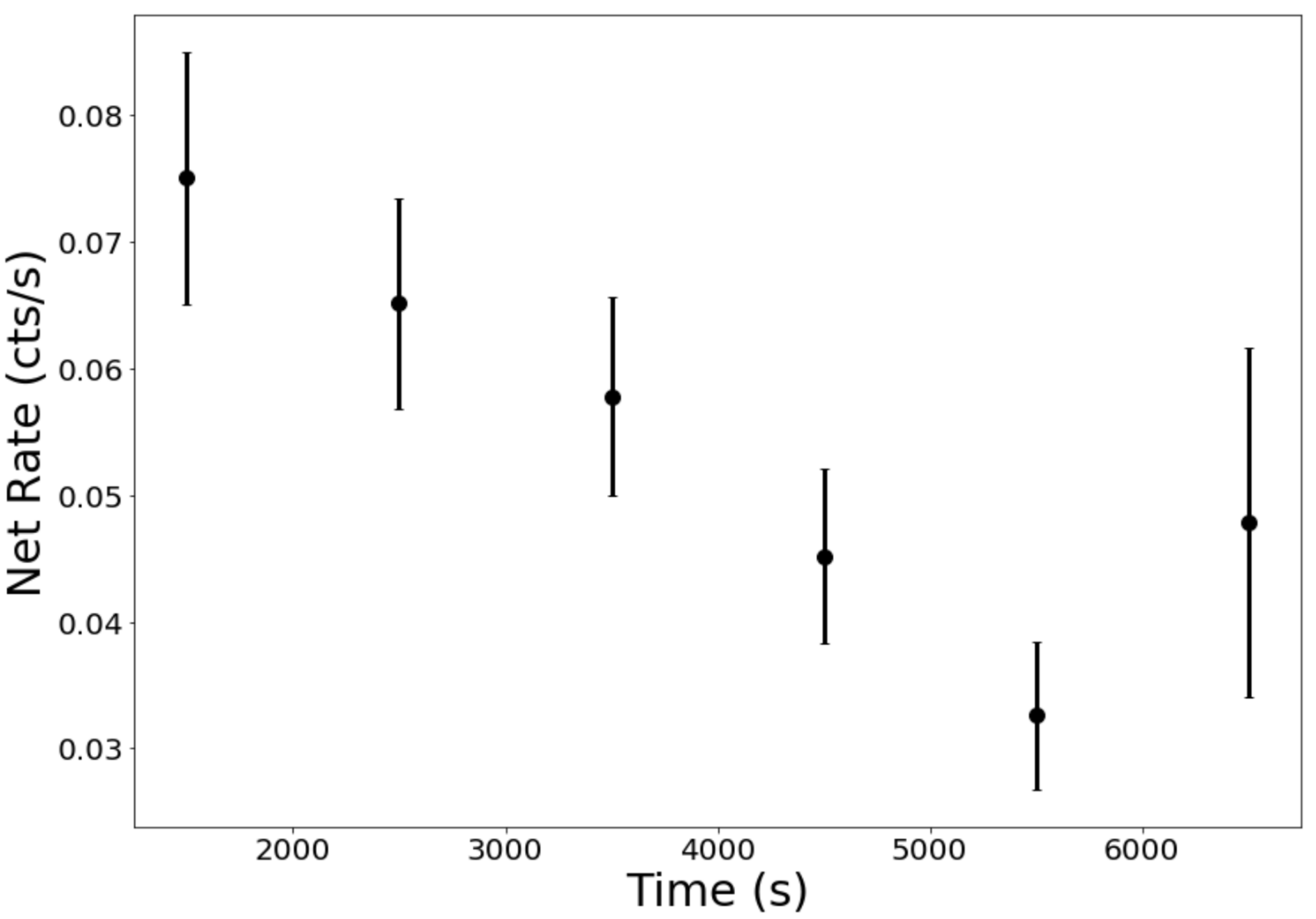}
\caption{{\sl Chandra} 0.5-8 keV net count rate during the 4.81 ks observation with a 1 ks binning. The source was variable on $\sim1$ ks timescales during this observation.
\label{cxo_cr}
}
\end{figure}

We have also searched for shorter timescale variability of  AX J1949 in the individual {\sl Chandra} and {\sl NuSTAR} observations. The 1 ks binned net count rate light curve from the {\sl Chandra} observation is shown in Figure \ref{cxo_cr}. The light curve clearly shows that the source brightness decreased throughout the observation on a several ks timescale. No significant flaring behavior was found in the {\sl Chandra} light curves with smaller binnings. We also searched for any periodic signal in the {\sl Chandra} data using the Z$^{2}_{1}$ \citep{1983A&A...128..245B}, but no significant periodicity was found.

\begin{figure}
\centering
\includegraphics[trim={0 0 0 0},scale=0.37]{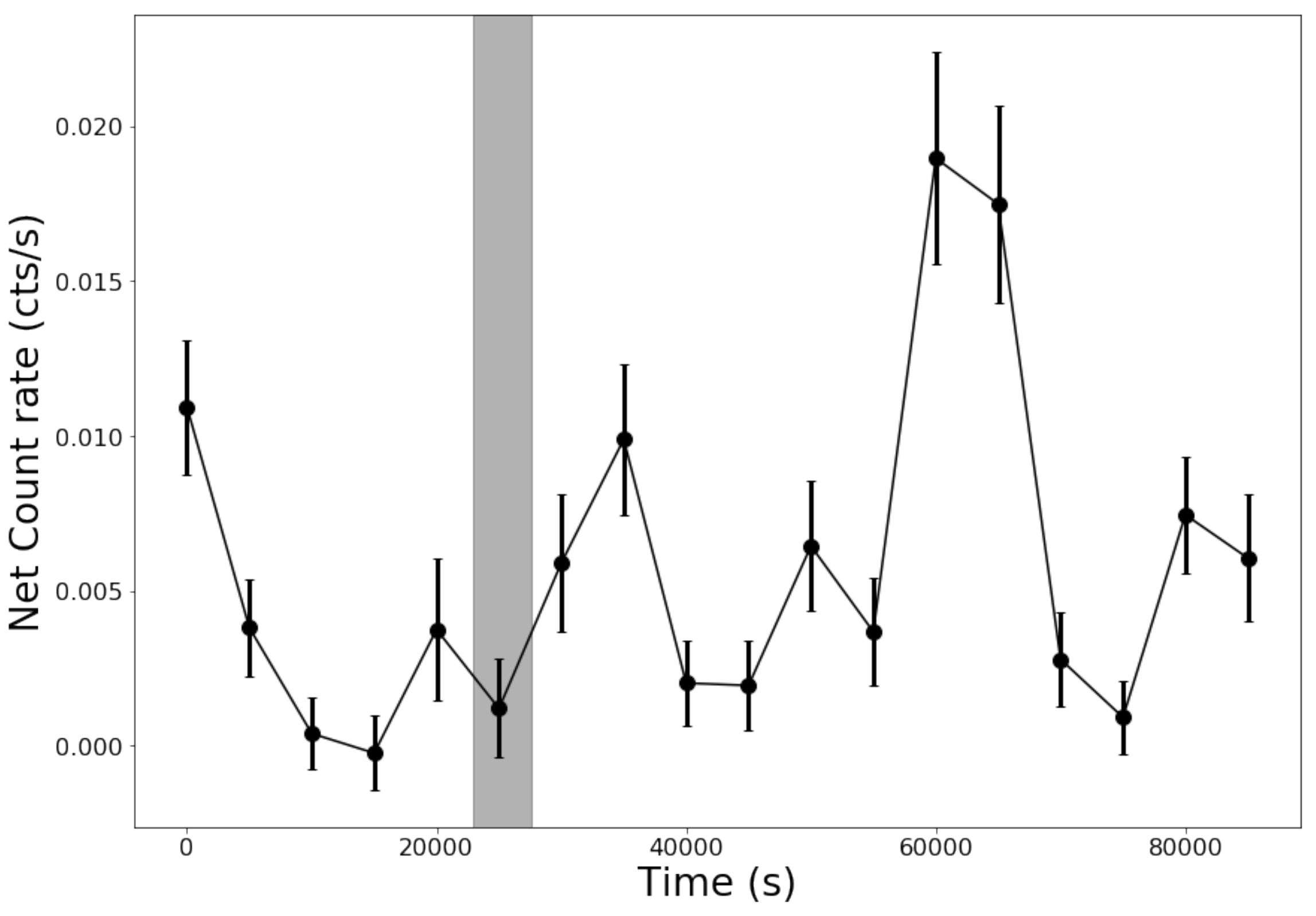}
\caption{{\sl NuSTAR} 3-20 keV  averaged (over the FPMA and FPMB detectors) net count rate light curve with a 5 ks binning. The source was variable on $\sim5$ ks timescales during this observation, including a  period of flux enhancement lasting $\sim10$ ks. The gray band shows the time and duration of the joint {\sl Swift-XRT} observation (observation S4). 
\label{nus_cr}
}
\end{figure}

\begin{figure}
\centering
\includegraphics[trim={0 0 0 0},scale=0.37]{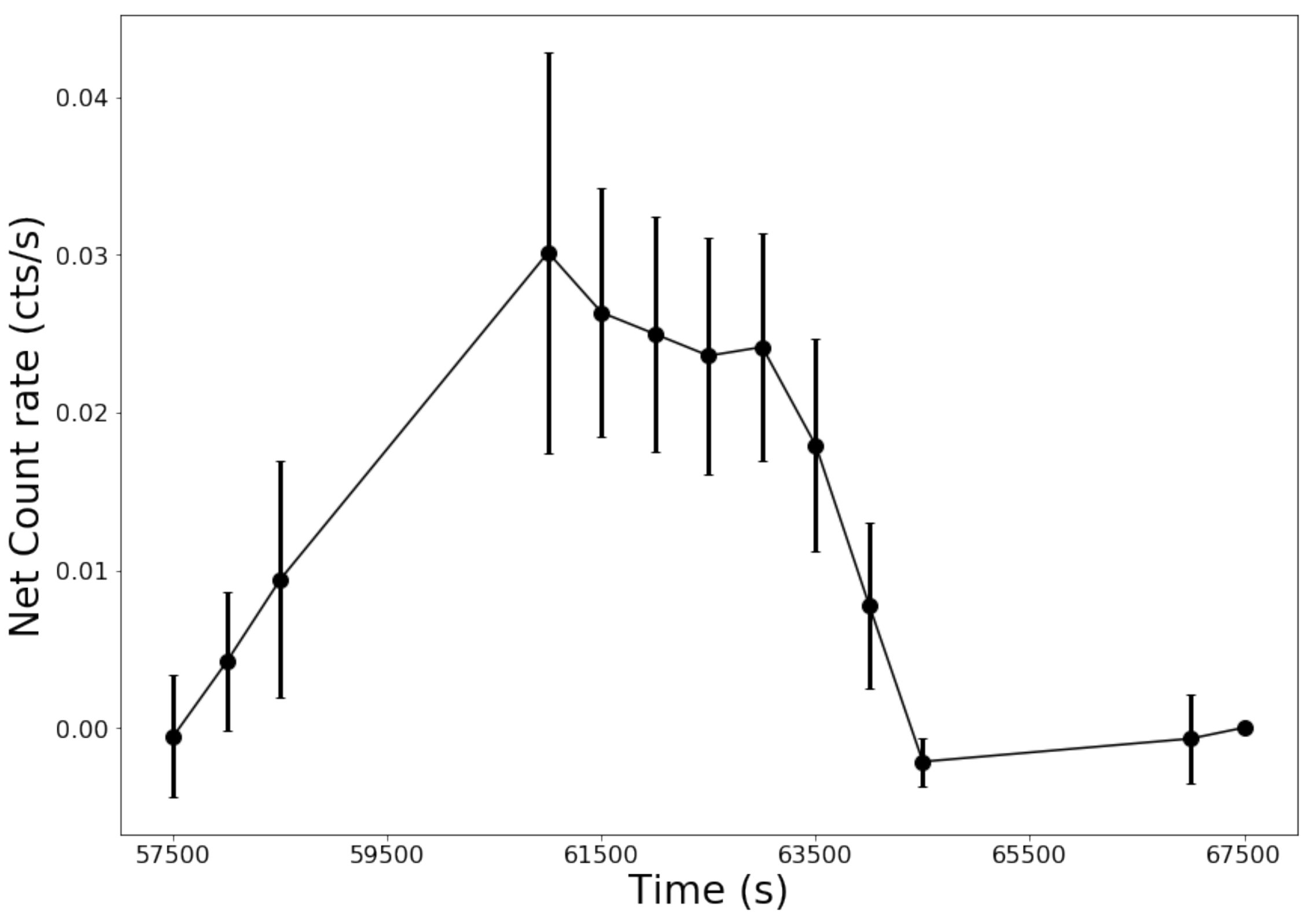}
\caption{{\sl NuSTAR} 3-20 keV net count rate light curve with a 500 s binning  during the period of flux enhancement. The source reaches a peak net count-rate of  $\sim$0.03 cts s$^{-1}$. 
\label{ns_flare}
}
\end{figure}

There  are also  indications of variability in the 3-20 keV {\sl NuSTAR} light curve. Notably, the 5 ks binned net light curve (see Figure \ref{nus_cr}) shows evidence of  a flux enhancement lasting $\sim$10 ks.   To test the significance of this variability we have fit a constant value to the light curve. The best-fit constant has a value of 3.0$\pm0.9\times10^{-3}$ counts s$^{-1}$ with a $\chi^2$=93.6 for 17 degrees of freedom, implying that the source is variable at a $\gtrsim6.5\sigma$ level. The 500 s binned light curve of the  period of flux enhancement (see Figure \ref{ns_flare}) shows the light curve reached a peak count rate of  $\sim$0.03 cts s$^{-1}$. This corresponds to a   peak flux of $\sim1.5\times10^{-12}$ erg cm$^{-2}$ s$^{-1}$, assuming the best-fit power law model (i.e., $\Gamma=1.2$ and $N_{H}=5.6\times10^{22}$ cm$^{-2}$). This flux is consistent with the flux observed during the {\sl Chandra} observation, suggesting that the source's  light curve varies over timescales of $\sim1-10$ ks.

\subsection{Optical/NIR companion photometry}

Using the accurate source position provided by {\sl Chandra} we have confidently identified the multi-wavelength counterpart to this source. Only one {\sl Gaia} source lies within the $\sim1''$ (2$\sigma$) positional uncertainty of  AX J1949. The source is also detected in optical by Pan-STARRs \citep{2016arXiv161205243F}, in NIR by 2MASS \citep{2006AJ....131.1163S}, and in IR by {\sl WISE} \citep{2012wise.rept....1C}. The counterparts multi-wavelength magnitudes can be found in Table \ref{mw_mag}.

\begin{table}[t!]
\caption{Multi-wavelength magnitudes of the counterpart to  AX J19498.} 
\label{mw_mag}
\small
\begin{center}
\renewcommand{\tabcolsep}{0.07cm}
\begin{tabular}{lcccc}
\tableline 
Obs. & R.A. & Decl. &  Filter &  Mag.  \\
\tableline 
 & deg. & deg. & &\\
\tableline
{\sl Gaia} & 297.480949745(9) &  25.56659555(1)  & G & 13.9860(8) \\
 & -- & --  & Bp & 16.299(6) \\
 & -- & --  & Rp & 12.578(3) \\
PS\tablenotemark{a} & 297.480960(2) & 25.566612(7)  & g & 17.516(8) \\ 
 & --  & --  & r & 14.981(4) \\
  & -- & --   & i & 13.1990 \\
  & --  & --   & z & 12.1140 \\
    & --  & --  & y & 11.7220 \\
    NOMAD & 297.4809331 & 25.5666150  & B & 18.04 \\
    & -- & -- & V & 16.12 \\
    & -- & -- & R & 14.62 \\
    2MASS & 297.480989 & 25.566639  & J & 9.90(2) \\
     & -- & -- & H & 9.07(2) \\
      & -- & -- & $K_s$ & 8.63(2) \\
      {\sl WISE} & 297.4809572 & 25.5666135  & W1 & 8.32(2) \\
      & --  & -- & $W2$ & 8.17(2) \\
            & -- & -- & $W3$ & 8.30(3) \\
            \tableline 
\end{tabular} 
\tablenotetext{a}{Pan-Starrs}
\end{center}
\end{table}

{\sl Gaia} has also measured the parallax, $\pi=0.081(64)$, and proper motion, $\mu_{\alpha}\cos{\delta}=-2.37(9)$, $\mu_{\delta}=-5.2(1)$, of  AX J1949's counterpart \citep{2018A&A...616A...1G}. Unfortunately, the {\sl Gaia} parallax has a large relative error ($\sim80\%$) and, therefore, has a large uncertainty in the inferred distance, $d=6.1^{+2.3}_{-1.5}$ kpc \citep{2018AJ....156...58B}. We note that this source also has a poorly fit astrometric solution with excess noise,  possibly due to its binary nature. Therefore, we do not rely on the {\sl Gaia} parallax and simply take it as an indicator that the source is at least at a distance of a few kpc. The large extinction observed in the optical magnitudes of the source also support this claim (see Section \ref{spec_type}).

\subsection{Optical spectroscopy}

The optical counterpart of  AX J1949 shows H$\alpha$ emission with 
EW$\approx -1.6$ \AA\ and FWZI $\approx500$ km~s$^{-1}$, and absorption
lines of \ion{He}{1} 5876 \AA\ and 6678 \AA\ (see Figure \ref{opt_spec}). The remaining absorption
features are diffuse interstellar bands, whose strengths
are consistent with the large extinction inferred from the colors
of the star (see Section \ref{spec_type}).

\section{Discussion}

\subsection{Optical/NIR Companion Stellar Type}
\label{spec_type}
The optical spectrum of the companion star shows He I absorption lines, suggesting that the star is hot (i.e., $\gtrsim$10,000 K) and, therefore, is likely to be of either O or B spectral class. Here we use the NOMAD $V$-band photometry in combination with the 2MASS NIR photometry to assess the spectral type, extinction, and distance to this star. The NOMAD $V$-band magnitude and the 2MASS NIR magnitudes can be found in Table \ref{mw_mag}. To estimate the extinction we use the intrinsic color relationships between the $V$-band and NIR magnitudes provided by Table 2 in \cite{2014AcA....64..261W}. To recover the intrinsic color relationships of late O and early to late B type stars, we need to deredden the source by an $A_V\approx8.5-9.5$, assuming the extinction law of \cite{1989ApJ...345..245C}. The 3D extinction maps\footnote{We use the {\tt mwdust} python package to examine the dust maps \citep{2016ApJ...818..130B}.}  of \cite{2006A&A...453..635M} suggest a distance of $\approx$ 7-8 kpc for this range of $A_V$.

The reddening and estimated distance to the source can be used to place constraints on its spectral type and luminosity class. In order for the star to have an observed $V$-band magnitude of 16.12 at a distance of 7 kpc with a reddening of $A_V=8.5$, it should have an absolute magnitude of $M_V\approx-6.6$. The lack of He II absorption features in the optical spectrum suggests that the star is more likely to be of B-type rather than O-type. By comparing the absolute magnitude to Table 7 in \cite{2006MNRAS.371..185W} we find that the star is most likely to be of the Ia luminosity class. Further, the strong He I absorption lines indicate that the star is an early B-type star. However, we mention one caveat, which is that while the lower luminosity classes are not bright enough to be consistent with the distance/reddening to the source, there is a large uncertainty in some of the absolute magnitudes for different luminosity classes (e.g., luminosity class Ib, II;  \citealt{2006MNRAS.371..185W}).

Lastly, the optical spectrum of the star shows an H$\alpha$ emission line, which is seen in both Be, and supergiant type stars. This line has been observed in a number of similar systems and is often variable in shape and intensity, sometimes showing a P-Cygni like profile (see e.g., \citealt{2006ApJ...638..982N,2006ESASP.604..165N,2006A&A...455..653P}). Unfortunately, we have only obtained a single spectrum and cannot assess the variability of the H$\alpha$ line. Additionally, it does not appear to be P-Cygni like in this single spectrum.

In comparison, by using optical and NIR photometry, \cite{2017MNRAS.469.3901S} found the stellar companion of  AX J1949 to be consistent with a B0.5Ia type star at a distance, $d=8.8$ kpc, and having a reddening, $A_V=7.2$. We have also found that the star is consistent with an early B-type type star of the Ia luminosity class. However, our estimates of the reddening and distance differ slightly from \cite{2017MNRAS.469.3901S}, in that we find a larger reddening and smaller distance.

\subsection{SFXT Nature}

At an assumed distance of $\sim7$ kpc,  AX J1949's {\sl Chandra} and {\sl Swift-}XRT fluxes (when the source is significantly detected) correspond to luminosities of $\sim1\times10^{34}$ erg s$^{-1}$ and $\sim3-8\times10^{33}$ erg s$^{-1}$, respectively.  Further, at harder X-ray energies,  AX J1949 has a  {\sl NuSTAR} 3$\sigma$ upper-limit quiescent luminosity $\lesssim 3\times10^{33}$ erg s$^{-1}$ in the 22-60 keV band, while during its flaring X-ray activity detected by {\sl INTEGRAL}, it reached a peak luminosity of $\sim10^{37}$ erg s$^{-1}$  in the 22-60 keV band \citep{2017MNRAS.469.3901S}. This implies a lower-limit on the dynamical range of the system of $\gtrsim 3000$. The large dynamical range and relatively low persistent X-ray luminosity are typical of SFXTs. Additionally, converting the absorbing column density, $N_{\rm H}=5.6\times10^{22}$ cm$^{-2}$, derived from the spectral fits to the X-ray data to $A_V$ following the relationship provided by \cite{2015MNRAS.452.3475B}, we find $A_V\approx20$. This $A_V$ is a factor of 2-2.5 larger than the $A_V$ derived from the interstellar absorption to the companion star (see Section \ref{spec_type}). This suggests that a large fraction of the absorption is intrinsic to the source, which has also been observed in other SFXTs (e.g., XTE J1739-302, AX J1845.0-0433; \citealt{2006ApJ...638..982N,2009A&A...494.1013Z}). Lastly, the optical counterpart to  AX J1949 appears to be an early supergiant B-type star, which have been found in $\sim40\%$ of the SFXT systems (see Figure 2 in \citealt{2017mbhe.confE..52S}). NIR spectra and additional optical spectra should be undertaken to search for additional absorption features, as well as changes in the H$\alpha$ line profile to solidify the spectral type.


\section{Summary and Conclusion}

A large number of unidentified {\sl INTEGRAL} sources showing rapid hard X-ray variability have been classified as SFXTs. We have analyzed  {\sl Neil Gehrels Swift-XRT}, {\sl Chandra}, and {\sl NuSTAR} observations of the SFXT candidate  AX J1949, which shows variability on ks timescales. The superb angular resolution of {\sl Chandra} has allowed us to confirm the optical counterpart to  AX J1949 and obtain its optical spectrum, which showed an H$\alpha$ emission line, along with He I absorption features. The spectrum, coupled with multi-wavelength photometry allowed us to place constraints on the reddening, $A_V=8.5-9.5$, and distance, $d\approx7-8$ kpc, to the source. We find that an early B-type Ia is the most likely spectral type and luminosity class of the star, making  AX J1949 a new confirmed member of the SFXT class. 

\medskip\noindent{\bf Acknowledgments:}
We thank Justin Rupert for obtaining the optical spectrum at MDM.  We thank the anonymous referee for providing useful and constructive comments that helped to improve the paper. This {\sl work} made use of observations obtained at the MDM Observatory, operated by Dartmouth College, Columbia University, Ohio State University, Ohio University, and the University of Michigan. This work made use of data from the {\it NuSTAR} mission, a project led by the California Institute of Technology, managed by the Jet Propulsion Laboratory, and funded by the National Aeronautics and Space Administration. We thank the {\it NuSTAR} Operations, Software and  Calibration teams for support with the execution and analysis of these observations.  This research has made use of the {\it NuSTAR}  Data Analysis Software (NuSTARDAS) jointly developed by the ASI Science Data Center (ASDC, Italy) and the California Institute of Technology (USA).  JH and JT acknowledge partial support from NASA through Caltech subcontract CIT-44a-1085101. JAT acknowledges partial support from Chandra grant GO8-19030X.

 \software{CIAO (v4.10; \citealt{2006SPIE.6270E..1VF}), XSPEC (v12.10.1; \citealt{1996ASPC..101...17A}), NuSTARDAS (v1.8.0), Matplotlib \citep{2007CSE.....9...90H}, Xselect (v2.4e), XIMAGE (v4.5.1), HEASOFT (v6.25), MWDust \citep{2016ApJ...818..130B}}


\begin{thebibliography}{}

\bibitem[Arnaud(1996)]{1996ASPC..101...17A} Arnaud, K.~A.\ 1996, Astronomical Data Analysis Software and Systems V, 101, 17
\bibitem[Bailer-Jones et al.(2018)]{2018AJ....156...58B} Bailer-Jones, C.~A.~L., Rybizki, J., Fouesneau, M., Mantelet, G., \& Andrae, R.\ 2018, \aj, 156, 58 
\bibitem[Bahramian et al.(2015)]{2015MNRAS.452.3475B} Bahramian, A., Heinke, C.~O., Degenaar, N., et al.\ 2015, \mnras, 452, 3475 
\bibitem[Barziv et al.(2001)]{2001A&A...377..925B} Barziv, O., Kaper, L., Van Kerkwijk, M.~H., et al.\ 2001, \aap, 377, 925.
\bibitem[Bovy et al.(2016)]{2016ApJ...818..130B} Bovy, J., Rix, H.-W., Green, G.~M., et al.\ 2016, \apj, 818, 130.
\bibitem[Bozzo et al.(2008)]{2008ApJ...683.1031B} Bozzo, E., Falanga, M., \& Stella, L.\ 2008, \apj, 683, 1031 
\bibitem[Bozzo et al.(2016)]{2016A&A...596A..16B} Bozzo, E., Bhalerao, V., Pradhan, P., et al.\ 2016, \aap, 596, A16 
\bibitem[Bozzo et al.(2016)]{2016A&A...589A.102B} Bozzo, E., Oskinova, L., Feldmeier, A., \& Falanga, M.\ 2016, \aap, 589, A102  
\bibitem[Buccheri et al.(1983)]{1983A&A...128..245B} Buccheri, R., Bennett, K., Bignami, G.~F., et al.\ 1983, \aap, 128, 245 
\bibitem[Burrows et al.(2005)]{2005SSRv..120..165B} Burrows, D.~N., Hill, J.~E., Nousek, J.~A., et al.\ 2005, \ssr, 120, 165 
\bibitem[Cardelli et al.(1989)]{1989ApJ...345..245C} Cardelli, J.~A., Clayton, G.~C., \& Mathis, J.~S.\ 1989, \apj, 345, 245 
\bibitem[Cash(1979)]{1979ApJ...228..939C} Cash, W.\ 1979, \apj, 228, 939 
\bibitem[Chaty et al.(2008)]{2008A&A...484..783C} Chaty, S., Rahoui, F., Foellmi, C., et al.\ 2008, \aap, 484, 783 
\bibitem[Chaty(2011)]{2011ASPC..447...29C} Chaty, S.\ 2011, Evolution of Compact Binaries, 447, 29 
\bibitem[Chaty(2015)]{2015arXiv151007681C} Chaty, S.\ 2015, arXiv:1510.07681 
\bibitem[Cutri et al.(2012)]{2012wise.rept....1C} Cutri, R.~M., Wright, E.~L., Conrow, T., et al.\ 2012, Explanatory Supplement to the WISE All-Sky Data Release Products,  
\bibitem[Ducci et al.(2009)]{2009MNRAS.398.2152D} Ducci, L., Sidoli, L., Mereghetti, S., Paizis, A., \& Romano, P.\ 2009, \mnras, 398, 2152 
\bibitem[Ducci et al.(2013)]{2013A&A...556A..72D} Ducci, L., Romano, P., Esposito, P., et al.\ 2013, \aap, 556, A72 
\bibitem[Evans et al.(2009)]{2009MNRAS.397.1177E} Evans, P.~A., Beardmore, A.~P., Page, K.~L., et al.\ 2009, \mnras, 397, 1177 
\bibitem[Filliatre \& Chaty(2004)]{2004ApJ...616..469F} Filliatre, P., \& Chaty, S.\ 2004, \apj, 616, 469 
\bibitem[Flewelling et al.(2016)]{2016arXiv161205243F} Flewelling, H.~A., Magnier, E.~A., Chambers, K.~C., et al.\ 2016, ArXiv e-prints , arXiv:1612.05243.
 \bibitem[Fruscione et al.(2006)]{2006SPIE.6270E..1VF} Fruscione, A., McDowell, J.~C., Allen, G.~E., et al.\ 2006, \procspie, 6270, 62701V 
\bibitem[Gaia Collaboration et al.(2018)]{2018A&A...616A...1G} Gaia Collaboration, Brown, A.~G.~A., Vallenari, A., et al.\ 2018, \aap, 616, A1 
\bibitem[Garmire et al.(2003)]{2003SPIE.4851...28G} Garmire, G.~P., Bautz, M.~W., Ford, P.~G., Nousek, J.~A., \& Ricker, G.~R., Jr.\ 2003, \procspie, 4851, 28 
\bibitem[Gehrels et al.(2004)]{2004ApJ...611.1005G} Gehrels, N., Chincarini, G., Giommi, P., et al.\ 2004, \apj, 611, 1005 
\bibitem[Goossens et al.(2018)]{2018arXiv181111882G} Goossens, M.~E., Bird, A.~J., Hill, A.~B., Sguera, V., \& Drave, S.~P.\ 2018, arXiv:1811.11882 
\bibitem[Grebenev \& Sunyaev(2007)]{2007AstL...33..149G} Grebenev, S.~A., \& Sunyaev, R.~A.\ 2007, Astronomy Letters, 33, 149 
\bibitem[Harrison et al.(2013)]{2013ApJ...770..103H} Harrison, F.~A., Craig, W.~W., Christensen, F.~E., et al.\ 2013, \apj, 770, 103 
 \bibitem[Hunter(2007)]{2007CSE.....9...90H} Hunter, J.~D.\ 2007, Computing in Science and Engineering, 9, 90 
\bibitem[in't Zand(2005)]{2005A&A...441L...1I} in't Zand, J.~J.~M.\ 2005, \aap, 441, L1 
\bibitem[Kim et al.(2007)]{2007ApJS..169..401K} Kim, M., Kim, D.-W., Wilkes, B.~J., et al.\ 2007, \apjs, 169, 401 
\bibitem[Krivonos et al.(2017)]{2017MNRAS.470..512K} Krivonos, R.~A., Tsygankov, S.~S., Mereminskiy, I.~A., et al.\ 2017, \mnras, 470, 512 
\bibitem[Lorenzo et al.(2010)]{2010ASPC..422..259L} Lorenzo, J., Negueruela, I., \& Norton, A.~J.\ 2010, High Energy Phenomena in Massive Stars, 422, 259 
\bibitem[Marshall et al.(2006)]{2006A&A...453..635M} Marshall, D.~J., Robin, A.~C., Reyl{\'e}, C., Schultheis, M., \& Picaud, S.\ 2006, \aap, 453, 635 
\bibitem[Matt \& Guainazzi(2003)]{2003MNRAS.341L..13M} Matt, G., \& Guainazzi, M.\ 2003, \mnras, 341, L13 
\bibitem[Negueruela et al.(2006a)]{2006ApJ...638..982N} Negueruela, I., Smith, D.~M., Harrison, T.~E., \& Torrej{\'o}n, J.~M.\ 2006, \apj, 638, 982 
\bibitem[Negueruela et al.(2006b)]{2006ESASP.604..165N} Negueruela, I., Smith, D.~M., Reig, P., Chaty, S., \& Torrej{\'o}n, J.~M.\ 2006, The X-ray Universe 2005, 604, 165 
\bibitem[Pellizza et al.(2006)]{2006A&A...455..653P} Pellizza, L.~J., Chaty, S., \& Negueruela, I.\ 2006, \aap, 455, 653 
\bibitem[Porter \& Rivinius(2003)]{2003PASP..115.1153P} Porter, J.~M., \& Rivinius, T.\ 2003, \pasp, 115, 1153 
\bibitem[Reig(2011)]{2011Ap&SS.332....1R} Reig, P.\ 2011, \apss, 332, 1 
\bibitem[Rivinius et al.(2013)]{2013A&ARv..21...69R} Rivinius, T., Carciofi, A.~C., \& Martayan, C.\ 2013, \aapr, 21, 69 
\bibitem[Romano et al.(2009)]{2009MNRAS.399.2021R} Romano, P., Sidoli, L., Cusumano, G., et al.\ 2009, \mnras, 399, 2021 
 \bibitem[Romano et al.(2011)]{2011MNRAS.412L..30R} Romano, P., Mangano, V., Cusumano, G., et al.\ 2011, \mnras, 412, L30.
\bibitem[Romano et al.(2014)]{2014A&A...568A..55R} Romano, P., Ducci, L., Mangano, V., et al.\ 2014, \aap, 568, A55 
\bibitem[Romano et al.(2015)]{2015A&A...576L...4R} Romano, P., Bozzo, E., Mangano, V., et al.\ 2015, \aap, 576, L4 
\bibitem[Sguera et al.(2005)]{2005A&A...444..221S} Sguera, V., Barlow, E.~J., Bird, A.~J., et al.\ 2005, \aap, 444, 221 
 \bibitem[Sguera et al.(2008)]{2008A&A...487..619S} Sguera, V., Bassani, L., Landi, R., et al.\ 2008, \aap, 487, 619 
\bibitem[Sguera et al.(2011)]{2011MNRAS.417..573S} Sguera, V., Drave, S.~P., Bird, A.~J., et al.\ 2011, \mnras, 417, 573 
\bibitem[Sguera et al.(2015)]{2015ATel.8250....1S} Sguera, V., Bazzano, A., \& Sidoli, L.\ 2015, The Astronomer's Telegram, 8250
\bibitem[Sguera et al.(2017)]{2017MNRAS.469.3901S} Sguera, V., Sidoli, L., Paizis, A., et al.\ 2017, \mnras, 469, 3901
\bibitem[Shakura et al.(2012)]{2012MNRAS.420..216S} Shakura, N., Postnov, K., Kochetkova, A., \& Hjalmarsdotter, L.\ 2012, \mnras, 420, 216 
\bibitem[Sidoli et al.(2009)]{2009MNRAS.397.1528S} Sidoli, L., Romano, P., Ducci, L., et al.\ 2009, \mnras, 397, 1528 
\bibitem[Sidoli(2013)]{2013arXiv1301.7574S} Sidoli, L.\ 2013, arXiv:1301.7574 
\bibitem[Sidoli(2017)]{2017mbhe.confE..52S} Sidoli, L.\ 2017, Proceedings of the XII Multifrequency 
Behaviour of High Energy Cosmic Sources Workshop.~12-17 June, 2017 Palermo, Italy
 \bibitem[Sidoli et al.(2017)]{2017ApJ...838..133S} Sidoli, L., Tiengo, A., Paizis, A., et al.\ 2017, \apj, 838, 133.
\bibitem[Sidoli \& Paizis(2018)]{2018MNRAS.481.2779S} Sidoli, L., \& Paizis, A.\ 2018, \mnras, 481, 2779 
\bibitem[Skrutskie et al.(2006)]{2006AJ....131.1163S} Skrutskie, M.~F., Cutri, R.~M., Stiening, R., et al.\ 2006, \aj, 131, 1163 
\bibitem[Sugizaki et al.(2001)]{2001ApJS..134...77S} Sugizaki, M., Mitsuda, K., Kaneda, H., et al.\ 2001, \apjs, 134, 77 
\bibitem[Walter et al.(2003)]{2003A&A...411L.427W} Walter, R., Rodriguez, J., Foschini, L., et al.\ 2003, \aap, 411, L427 
\bibitem[Walter \& Zurita Heras(2007)]{2007A&A...476..335W} Walter, R., \& Zurita Heras, J.\ 2007, \aap, 476, 335 
\bibitem[Walter et al.(2015)]{2015A&ARv..23....2W} Walter, R., Lutovinov, A.~A., Bozzo, E., \& Tsygankov, S.~S.\ 2015, \aapr, 23, 2  
\bibitem[Wegner(2006)]{2006MNRAS.371..185W} Wegner, W.\ 2006, \mnras, 371, 185 
\bibitem[Wegner(2014)]{2014AcA....64..261W} Wegner, W.\ 2014, \actaa, 64, 261 
\bibitem[Wegner(2015)]{2015AN....336..159W} Wegner, W.\ 2015, Astronomische Nachrichten, 336, 159 
\bibitem[Wilms et al.(2000)]{2000ApJ...542..914W} Wilms, J., Allen, A., \& McCray, R.\ 2000, \apj, 542, 914 
\bibitem[Winkler et al.(2003)]{2003A&A...411L...1W} Winkler, C., Courvoisier, T.~J.-L., Di Cocco, G., et al.\ 2003, \aap, 411, L1 
\bibitem[Zurita Heras \& Walter(2009)]{2009A&A...494.1013Z} Zurita Heras, J.~A., \& Walter, R.\ 2009, \aap, 494, 1013 






\end{thebibliography}
\end{document}